# Performance of Packet-to-Cell Segmentation Schemes in Input Buffered Packet Switches

K. Christensen and K. Yoshigoe
Computer Science and Engineering
University of South Florida
Tampa, FL 33620

A. Roginsky
IBM Corporation
P.O. Box 12195
Research Triangle Park, NC 27709

N. Gunther
Performance Dynamics Company
4061 East Castro Blvd., Suite 110
Castro Valley, CA 94552

*Abstract*- **Most input buffered packet switches internally segment variable-length packets into fixed-length cells. The last cell in a segmented packet will contain overhead bytes if the packet length is not evenly divisible by the cell length. Switch speed-up is used to compensate for this overhead. In this paper, we develop an analytical model of a single-server queue where an input stream of packets is segmented into cells for service. Analytical models are developed for M/M/1, M/H$_2$/1, and M/E$_2$/1 queues with a discretized (or quantized) service time. These models and simulation using real packet traces are used to evaluate the effect of speed-up on mean queue length. We propose and evaluate a new method of segmenting a packet trailer and subsequent packet header into a single cell. This cell merging method reduces the required speed-up. No changes to switch-matrix scheduling algorithms are needed. Simulation with a packet trace shows a reduction in the needed speed-up for an iSLIP scheduled input buffered switch.**

## I. INTRODUCTION

Packet switching is central to IP routing and Ethernet switching. Packets are typically variable in length (e.g., ranging from 64 to 1518 bytes for Ethernet). To accommodate very high-speed links, input buffered architectures are often used (e.g., the Cisco 12000 GSR [1]). These input buffered switches can use a crossbar as the switch fabric. Input buffered switches require a memory speed proportional to link speed. Output buffered switches require memory speed proportional to $N$ times link speed (for $N$ ports) making them infeasible for use with high-speed links and/or large numbers of ports. Input buffered switches overcome head-of-line blocking by employing virtual output queueing (VOQ) in each input port. A VOQ input buffered switch contains $N$x$N$ VOQs for $N$ ports. Fig. 1 shows a VOQ switch with each input port containing a packet classifier to determine the destination output port, packet-to-cell segmenter, VOQs, and a scheduler. VOQ switches use an iterative matching algorithm to match input ports with output ports to schedule the switch matrix. Iterative matching algorithms for VOQ crossbar switches include iSLIP [2] and FIRM [3]. These algorithms inherently require the use of internal fixed-length cells. This is because the crossbar is scheduled in cycles, one cycle for each set of cells forwarded from matched input ports to output ports. Thus, input buffered switches segment packets into cells, internally switch the cells, and then reassemble the cells into packets in reassembly buffers at the output ports. "Cell train" approaches to reducing packet-level delays have been studied (e.g., [4]).

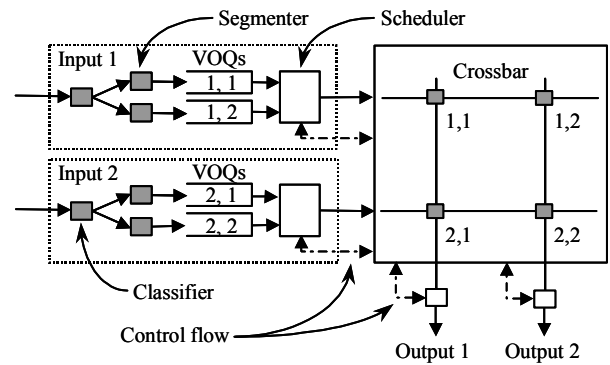

Fig. 1. VOQ switch showing packet-to-cell segmenter

Not all (very likely, only very few) packets have a length that is an even multiple of the internal cell length. Thus, the last cell of a segmented packet will often contain padding bytes. To compensate for internal forwarding of overhead bytes, the switch buffers and fabric must operate faster than link speed. That is, an internal speed-up is needed to achieve queue stability for high offered loads. In the worst case of $S$ byte cells and contiguous arrivals of $S+1$ byte packets, the speed-up needs to be almost a factor of 2 for $2S$ bytes of cell data needed to switch $S+1$ bytes of packet data. Speed-up is impossible when the memory and logic speeds of the switch technology are already close to being exceeded by link speeds. In any case, the overhead from speed-up adds cost to a switch implementation. No existing work quantitatively addresses how much speed-up is needed. New methods for reducing speed-up need to be investigated. We address these open problems.

This paper is organized as follows. Section II of this paper develops fundamental analytical models and contains numerical results from these models. Section III contains a simulation evaluation using real packet traces. Section IV proposes and evaluates a method of consecutive packets in cells to reduce overhead. Section V is a summary. An appendix contains proofs.

This material is based upon work supported by the National Science Foundation under Grant No. 9875177.

## II. MODELS OF QUANTIZED SERVICE TIME QUEUES

We assume that packet lengths are randomly distributed and their distribution is known. For a given packet of length $L$ bytes and a cell of size $S$ bytes, $\text{ceil}(L/S)$ cells are needed to segment the packet where ceil is the standard ceiling function. The last packet in the sequence, or "train", of cells will have $S-r$ padding or overhead bytes where $r = \text{Rem}(L,S)$ when $L$ is not a multiple of $S$, $r = S$ otherwise. This results in an internal packet length (in cells) of $L+S-r$ bytes. Thus, an internal speed-up factor of $1+(S-r)/L$ is needed to switch the packet at link speed. For a given distribution of packet lengths, we address how this discretization (or quantization) caused by segmentation into cells changes the mean and variance of the number of bytes to be transported. This changes the service time of a queue modeling the input buffer of a packet switch that segments packets to cells as described. We model queueing delay of three classical queues given a ceiling of service time.

### A. Ceiling of Well Known Distributions

Let $X$ be an arbitrary random variable for which we know $E(X)$ and $Var(X)$. Let $Y$ be the integer-valued random variable that is the ceiling of $X$ and let $Z = Y - X$. If $X$ has a wide smooth distribution and is not concentrated near a particular integer or set of integers, we can assume that $Y$ and $Z$ are almost independent and the distribution of $Z$ on $(0,1)$ is roughly uniform. We have $E(Z) = 1/2$ and $Var(Z) = 1/12$. We have $E(Y) = E(X+Z) = E(X) + E(Z) = E(X) + 1/2$ and $Var(X) = Var(Y-Z) = Var(Y) + Var(Z) = Var(Y) + 1/12$ and hence $Var(Y) = Var(X) - 1/12$.

For an exponentially distributed random variable $X$ with $f_X(x) = \mu e^{-\mu x}$, let $Y = \text{ceil}(X)$. Then $Y$ is geometrically distributed with:

$$E(Y) = \frac{1}{1-e^{-\mu}} \text{ and} \quad (1)$$

$$Var(Y) = \frac{e^{-\mu}}{(1-e^{-\mu})^2}. \quad (2)$$

The proof of this is in the appendix. As expected,

$$\lim_{\mu \to 0}\left(\frac{1}{1-e^{-\mu}} - \frac{1}{\mu}\right) = \frac{1}{2} \text{ and} \quad (3)$$

$$\lim_{\mu \to 0}\left(\frac{e^{-\mu}}{(1-e^{-\mu})^2} - \frac{1}{\mu^2}\right) = -\frac{1}{12}. \quad (4)$$

Let $H_2$ denote a two-stage hyperexponentially distributed random variable $X$ with $f_X(x) = \alpha_1 \mu_1 e^{-\mu_1 x} + \alpha_2 \mu_2 e^{-\mu_2 x}$ and $x > 0$, $\alpha_1, \alpha_2 \geq 0$, and $\alpha_1 + \alpha_2 = 1$, let $Y = \text{ceil}(X)$. Then,

$$E(Y) = \frac{\alpha_1}{1-e^{-\mu_1}} + \frac{\alpha_2}{1-e^{-\mu_2}} \text{ and} \quad (5)$$

$$Var(Y) = \alpha_1 \frac{1+e^{-\mu_1}}{(1-e^{-\mu_1})^2} + \alpha_2 \frac{1+e^{-\mu_2}}{(1-e^{-\mu_2})^2} - (E(Y))^2. \quad (6)$$

The proof of this is in the appendix.

Let $E_2$ denote a two-stage Erlang distributed random variable $X$ with $f(x) = \mu^2 x e^{-\mu x}$ (i.e., both service rates are the same). Then $Y$ has,

$$E(Y) = \frac{(\mu-1)e^{-\mu} + 1}{(1-e^{-\mu})^2} \text{ and} \quad (7)$$

$$Var(Y) = \frac{[(\mu+1)e^{-\mu} - (\mu-1)e^{-3\mu} - (\mu^2+2)e^{-2\mu}]}{(1-e^{-\mu})^4}. \quad (8)$$

The proof of this is in the appendix. When $Var(Y)$ is evaluated numerically, we can see that it is usually greater than the value of $Var(X)$, which is equal to $2/\mu^2$. The reason for this is that the Erlang distribution does not satisfy the heuristic assumptions made in Section IIA. It has a peak, which makes the discrete picture more complicated.

### B. M/G/1 Analysis

The quantization described in Section IIA corresponds to taking the ceiling function of the continuous-valued service time. We introduce a superscript notation $\Delta$ for these quantized service times so that the quantized version of an M/M/1 queue is denoted M/M$^\Delta$/1, which is also equivalent to M/Geo/1 with mean service time $(1-e^{-\mu})^{-1}$ defined by eq. (1). Since $E(Y)$ and $Var(Y)$ are continuous functions, the mean number of customers in the system $E(N)$ can be determined from the standard Pollaczek-Khintchine (P-K) M/G/1 formula [5]:

$$E(N) = \rho + \frac{\rho^2}{2(1-\rho)}\left[1 + \frac{Var(Y)}{E^2(Y)}\right]. \quad (9)$$

where $\rho = \lambda E(Y)$ is the quantized load. For M/M$^\Delta$/1 eq. (9) reduces to:

$$E(N) = \frac{\lambda}{2}\left(\frac{\lambda-2}{e^{-\mu} + \lambda - 1}\right) \quad (10)$$

with $\mu$ the mean service rate of the underlying exponential distribution used to calculate the quantized service times. We have also solved M/H$_2^\Delta$/1 and M/E$_2^\Delta$/1 queues for $E(N)$, but those formulas are far more complex than eq. (10) and are not shown here. Moreover, beyond the special cases discussed in Section IIA, we do not expect an arbitrary response time distribution to possess an analytic quantized form, so a more general approach is needed.

### C. General Quantization Algorithm

Let $f(t)$ be a continuous service time probability density, then the following algorithm can be applied either algebraically or numerically:

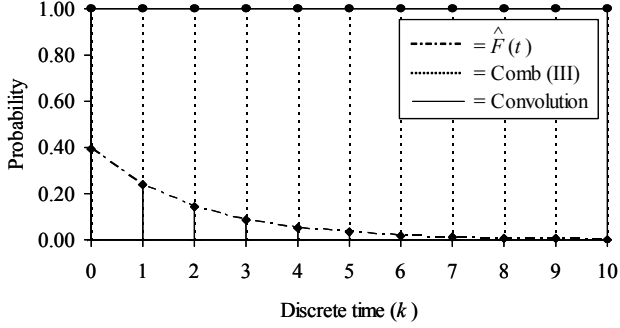

Fig 2. Dirac comb for $\hat{F}(t)$ (exponential with $\mu = 0.5$)

1. Determine the cumulative distribution function

$$F(t) = \int_0^t f(x)dx \quad (11)$$

which corresponds to the *total area* or probability.

2. Construct the *continuous* difference function:

$$\hat{F}(t) = F(t+1) - F(t). \quad (12)$$

which corresponds to the sub-area in each time-domain interval. The procedure, so far, is the same as that used for proofs in the Appendix. If $\hat{F}(t)$ does not possess a closed analytic form, numerical techniques can be employed.

3. Convolve eq. (12) with the *Dirac comb* III [9] to produce:

$$\hat{F}(t) \cdot III = \sum_{k=0}^{\infty} \hat{F}(k)\delta(t-k-1), \quad (13)$$

where $\delta(\cdot)$ is the Dirac delta function [7] chosen to conform to the conventions used in the Appendix. Selecting integer sampling points with $\Delta k = 1$ corresponds to applying the ceiling function of Section IIA (see Fig. 2). However, eq. (13) is guaranteed to be the correct discrete representation for any arbitrary continuous service time density $f(t)$.

4. The corresponding mean and variance of the quantized distribution are computed from:

$$E(Y) = \sum_{k=1}^{N} k\hat{F}(k)\delta(t-k-1) \text{ and} \quad (14)$$

$$Var(Y) = \sum_{k=1}^{N} (k - E_\Gamma(Y))^2 \hat{F}(k)\delta(t-k-1). \quad (15)$$

As an example applicaton of this quantization algorithm, consider an M/Γ/1 queue with Gamma probability density defined by:

$$f(t, \alpha, \beta) = \frac{t^{\alpha-1}}{\beta^\alpha \Gamma(\alpha)} e^{-t/\beta} \quad (16)$$

TABLE I
SUMMARY NETWORK STATISTICS (from [8])

| Statistic | Value |
|---|---|
| Sample mean ($m$) | 1.74 |
| Sample standard deviation ($s$) | 0.89 |
| Estimated shape ($\alpha$) | 3.82 |
| Estimated scale ($\beta$) | 0.46 |

where the gamma function $\Gamma(\alpha+1) = \alpha!$ is a generalization of the factorial function [7]. The parameters $\alpha$ and $\beta$ respectively determine the *shape* and the *scale* of eq. (16), which makes it attractive for modeling a variety of empirical service time distributions [8]. If $\alpha = 1$, eq. (16) reduces to the exponential density function discussed in Section IIA. Unfortunately, this inherent modeling flexibility comes with the limitation that no analytic form exists for the quantized distribution $\hat{F}_\Gamma(t) \cdot III$ in eq. (13). We therefore proceed numerically.

Using the summary network statistics in Table 1, we have $\alpha = (m/s)^2 = 3.82$ and $\beta = m/\alpha = 0.46$. Substituting these values into eqs. (14) and (15) produces $E_\Gamma(Y) = 2.24$ and $Var_\Gamma(Y) = 0.89$, respectively and eq. (9) gives the mean queue length of M/Γ$^\Delta$/1 (e.g., $E_\Gamma(N) = 5.67$ at an offered load of $\rho = 0.90$ (compare this with Fig. 3)).

Eq. (13) can also be used to construct the appropriate queue-theoretic transforms for the quantized service time distribution from the Laplace transform:

$$W[(1-z)\lambda] = \int_0^\infty \hat{F}(t) \cdot III \ e^{-(1-z)\lambda t} dt. \quad (17)$$

For the M/M$^\Delta$/1 queue:

$$W[(1-z)\lambda] = \frac{e^{-(1-z)\lambda}(1-e^\mu)}{e^{-(1-z)\lambda} - e^\mu} \quad (18)$$

and the corresponding P-K transform [5] of the queue length distribution:

$$g(z) = \frac{(1-\rho)(1-z)W[(1-z)\lambda]}{W[(1-z)\lambda] - z} \quad (19)$$

reduces to:

$$g(z) = \frac{(1-z)e^{-(1-z)\lambda}\left[(1-\lambda)e^\mu - 1\right]}{e^{-(1-z)\lambda+\mu} - (1-z)e^{-(1-z)\lambda} - ze^\mu}. \quad (20)$$

All the moments of the queue length distribution can be found by taking successive derivatives of $g(z)$ [5]. The first moment is the mean queue length:

$$E(N) \equiv \left.\frac{dg(z)}{dz}\right|_{z=1} = \frac{\lambda(\lambda-2)e^\mu}{2 - 2(1-\lambda)e^\mu}, \quad (21)$$

which is equivalent to eq. (10). The inverse of $g(z)$ (if it exists) produces the queue length probability distribution.

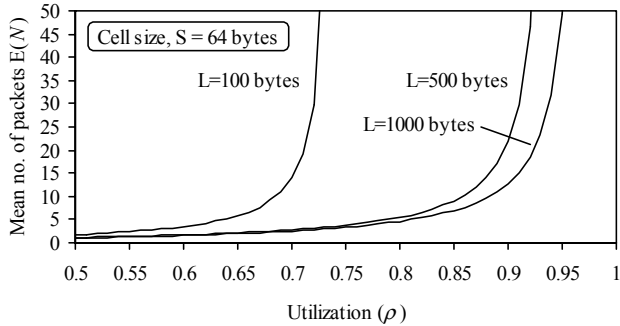

Fig 3. Numerical results for M/M$^\Delta$/1 for various values of *L*

## D. Application of Models to Packet-to-Cell Segmenting

We now apply the M/M$^\Delta$/1 model to predict $E(N)$ in packets given *L*, *S*, and a utilization $\rho$. The utilization, $\rho$, is based on the arriving rate of packets (in bits per second) divided by the link rate (in bits per second). Packet arrivals are Poisson, and packet lengths are exponentially distributed. These assumptions are very restrictive and unrealistic of real packet traffic. However, even with these restrictive assumptions we can observe the general behavior of segmentation and speed-up. In the next section, these restrictive assumptions are removed in a simulation study.

For a given *L* and *S*, the mean service time in cell time units is $T_s = L/S$. For a given link utilization based on packets, $\rho$, the mean interarrival time is $T_a = T_s/\rho$. Then, the mean arrival rate $\lambda = 1/T_a$ and mean service rate $\mu = 1/T_s$. For the M/M/1, $\rho = \lambda/\mu$ and mean number of packets in the system $E(N) = \rho/(1-\rho)$. For the M/M$^\Delta$/1 the utilization for quantized service time is, $\rho' = \lambda \cdot E(Y) = \lambda/(1 - e^{-\mu})$. The speed-up, $\sigma$, needed to achieve carried load equal to offered load (i.e., stability for all offered loads up to $\rho = 1$) is $\sigma = \rho'/\rho = \mu \cdot E(Y) = \mu/(1 - e^{-\mu})$. Fig. 3 shows the numerical results for $E(N)$ for a range of $\rho = 0.50, 0.51, \ldots, 1.0$ and $L = 100$, 500, and 1000 bytes for $S = 64$ bytes. To achieve stability, the speed-up required is $\sigma = 1.354$ for $L = 100$ bytes, $\sigma = 1.065$ for $L = 500$ bytes, and $\sigma = 1.032$ for $L = 1000$ bytes. These numerical results show that without speed-up, $E(N)$ increases rapidly at high offered loads and that the smaller *L* is, the greater is the effect of segmentation on $E(N)$. The analytical results in Fig. 3 have been validated with a simulation model. This model was then used in the next section for more realistic traffic models.

## III. SIMULATION OF iSLIP WITH PACKET SEGMENTATION

We evaluated the effects of discretization on a real network traffic using a simulation model with a packet trace as input. We used a previously built and validated iSLIP simulation model [6] for this simulation evaluation. Over 60 million IP packets were collected from the University of South Florida (USF) Internet2 OC-3 (155-Mbps) link. The packet trace collected packet interarrival times, packet length, and packet headers. Table 2 shows the summary of the packet trace. The mean packet length was 764 bytes. The most common packet length was 1518 bytes (31.4% of all packets) followed by 64 bytes (28.7%), 1438 bytes (7.7%), 70 bytes (2.7%), and 594 bytes (1.4%). All other packet length occurs at less than 1%. The trace file of 60 million packets was split into 16 smaller files, each with the same number of packets. Each of these smaller files was then input to a port in the modeled 16-port iSLIP switch. The destination output port was assigned using a modulo-16 function of the packet IP destination address. Output port utilization of switch is not uniform for real traffic. In our experiment, we refer utilization to be the maximum offered load among all 16 ports. Service time (i.e., simulated line speed) is controlled to achieve a desired utilization. For all experiments, control variables are offered load and speed-up, and the response variable is mean queue length. An internal cell size of 64 bytes was used. Two experiments were run:

TABLE II
PACKET TRACE SUMMARY STATISTICS

|  | Packet interarrival time | Packet length |
|---|---|---|
| Mean | 66.50 $\mu$s | 764 bytes |
| Standard deviation | 68.20 $\mu$s | 672 bytes |
| CoV | 1.03 | 0.88 |
| Minimum | 0.95 $\mu$s | 64 bytes |
| Maximum | 2385.97 $\mu$s | 1518 bytes |
| 99% | 338.08 $\mu$s | 1518 bytes |
| Shape of distribution | Decreasing | Bimodal |

*Experiment #1* – For no speedup, 1.05x and 1.1x speedups with segmentation and cell padding, mean queue length is measured for utilization ranging from 50% to 99%.

*Experiment #2* – The minimum speedup needed for 99% utilization is systematically identified. A queue length of 1000 or greater is considered a sign of instability.

Fig. 4 shows the results for the experiment #1. It can be seen that no speedup and the 1.05x speedup cases become unstable above 93% and 97% utilization, respectively. The 1.1x speedup case can achieve stability for the entire range of utilizations. For experiment #2, it was found that the minimum speed-up needed for 99% utilization was 1.06x.

## IV. PACKET-TO-CELL SEGMENTATION WITH CELL MERGING

We propose a new method of segmenting packets into cells that applies to input buffered switches and reduces the amount of speed-up needed to achieve stability. When a packet is segmented into cells (i.e., in the segmenters shown in Fig. 1) the last cell of a packet may be a partially filled cell. Instead of queueing this partial cell (with padding bytes) to the VOQ, it is held back to wait for the next arriving packet. The next arriving packet then starts its segmentation with the held back cell from the previous packet. That is, the header bytes of the arriving packet are merged with the trailer bytes from the previous packet. We call this cell merging. A finite state machine (FSM) for cell merging is shown in Fig. 5. The EMPTY state occurs when there is no held back cell and the segmenter is idle. An arriving packet transitions (T0) the FSM to the SEGMENTING state where the packet is segmented into cells and the cells queued in the VOQ. If the last cell in a packet is a partial cell it is held back and the FSM transitions (T1) to the PARTIAL state (otherwise, the transition is to the EMPTY state (T2)). In the PARTIAL state, the segmenter is idle and waiting for an arriving packet to transition (T3) back to the SEGMENTING state. In the PARTIAL state a cell merging timer is started when the VOQ is empty (e.g., all cells segmented and queued have been forwarded). If this timer expires before an arriving packet, then the held back cell is queued with padding bytes and transition (T4) is to the EMPTY state. The purpose of the timer is to prevent a packet from being unfairly starved if there are no subsequent arrivals for a long period of time.

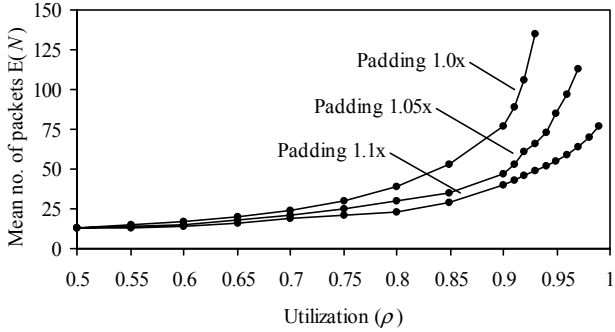

Fig 4. Simulation results for experiment #1

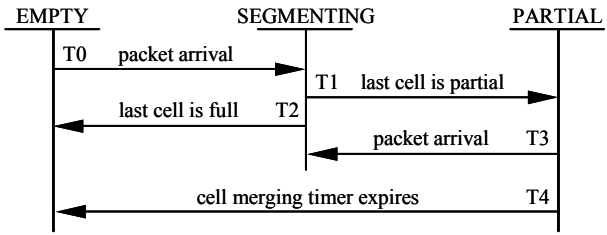

Fig. 5. FSM for cell merging

*A. Simulation Evaluation of Cell Merging*

Experiments #1 and #2 of Section III were repeated for the cell merging mechanism with the cell merging timer expiration value set to 10 cell times. Fig. 6 shows the mean queue of cell merging compared with the results of Fig. 4 (no cell merging). The packet merging mechanism with no speedup becomes unstable above 95%. The cell merging mechanism with a speedup of 1.05x and 1.1x achieved stability for all offered load measured. Cell merging results in a lower mean queueing delay for high utilizations. From experiment #2 it was found that the minimum speedup needed was 1.04x. Thus, cell merging required 2% less speedup than packet-to-cell segmentation without cell merging. We experimented with cell merging timer values. A large (100 or 1000 cell times) timer value results in high queueing delays at low utilization and no benefit at high utilizations (where a time-out would rarely occur due to frequently arriving packets). We found that a value of 10 cells times works well.

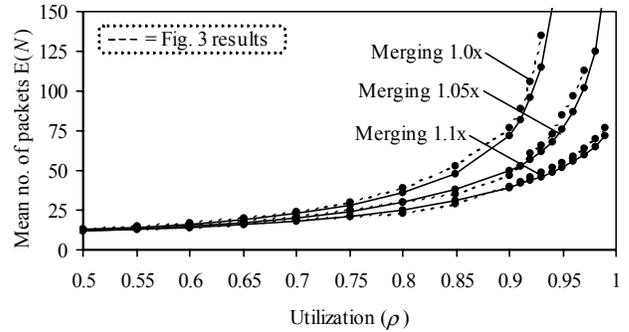

Fig 6. Simulation results for experiment #1 with cell merging

For a 10-Gbps link, 10 cell times corresponds to a very small 512 ns.

V. SUMMARY AND FUTURE WORK

This paper makes two contributions. The first is in discovering the application of quantized queues to modeling of packets-to-cells segmentation for packet switches. This is the first work to analytically show the speed-up value needed for a packet-to-cell segmentation. We have formally derived and proved expressions for mean queue length for $M/M^{\Delta}/1$, $M/H_2^{\Delta}/1$, and $M/E_2^{\Delta}/1$. We have also described a general quantization procedure. Our second contribution is to evaluate the speed-up penalty of packets-to-cells segmentation and propose an improved method of segmentation. Our proposed cell merging method reduces the required speed-up in an input buffered switch. This is significant because increases in link speed continue to outpace improvements in memory speed. We cannot afford up to a 2x speed-up just for handling the variable length nature of IP packets in the Internet. Even a small speed-up adds cost to high-speed packet switches.

We believe that quantized queues are an area of much future work in both theory and application. Future work includes investigating inverting the P-K transform to determine mean queue length distributions. Future work also includes investigating additional practical methods to further reduce the required speed-up in VOQ switches.

ACKNOWLEDGMENT

The authors thank the anonymous referees for their helpful comments that have improved the quality of this paper. The authors also thank Bjarne Helvik of NTNU for his discussion on the properties of the distribution of *Z* in Section IIA

REFERENCES

[1] *Cisco 12000 Gigabit Switch Router* [Online]. Available: http://www.cisco.com.
[2] N. McKeown, "The iSLIP Scheduling Algorithm for Input-Queued Switches," *IEEE/ACM Trans. Networking*, vol. 7, no. 2, Apr. 1999, pp. 188-201.
[3] D. Serpanos and P. Antoniadis, "FIRM: A Class of Distributed Scheduling Algorithms for High-Speed ATM Switches with Multiple Input Queues," *Proc. IEEE INFOCOM*, Mar. 2000, pp. 548-55.

[4] M. Marsan, A. Bianco, P. Giaccone, E. Leonardi, and F. Neri, "Packet-Mode Scheduling in Input-Queued Cell-Based Switch," *IEEE/ACM Trans. Networking*, vol. 10, no. 5, Oct. 2002, pp. 666-77.
[5] L. Kleinrock, *Queueing Systems, Volume 1: Theory*, New York: John Wiley & Sons, 1975.
[6] K. Yoshigoe and K. Christensen, "An Evolution to Crossbar Switches with Buffered Cross Points," *IEEE Network*, vol. 7, no. 5, Sept.-Oct. 2003, pp. 48-56.
[7] M. Abramowitz and A. Stegun, *Handbook of Mathematical Functions*, New York: Dover Press, 1970.
[8] N. Gunther, *The Practical Performance Analyst*, New York: McGraw-Hill, 1998.
[9] A. Oppenheim, A Willsky, and I. Young, *Signals and Systems*, New Jersey: Prentice-Hall, 1983.

APPENDIX

In the appendix we derive formulas for the mean and variance of the ceiling of exponentially, hyperexponentially, and two-stage Erlang distributed random variables. That is, proofs are given for eq. (1) and (2) for exponential, eq. (5) and (6) for hyperexponential, and eq. (7) and (8) for two-stage Erlang.

Proof of eq. (1) and (2): For any $k \geq 1$ we compute

$$P(Y = k) = P(k-1 < X \leq k) =$$

$$F_X(k) - F_X(k-1) = \left(1 - e^{-\lambda k}\right) - \left(1 - e^{-\lambda(k-1)}\right). \quad (A1)$$

We can rewrite this $P(Y = k) = e^{-\lambda(k-1)}\left(1 - e^{-\lambda}\right)$. Let $Z = Y - 1$. Then $P(Z = k) = P(Y = k+1) = e^{-\lambda k}\left(1 - e^{-\lambda}\right)$. From this it follows that $Z$ has a geometric distribution, that is, $P(Z = k) = q^k p$, $p + q = 1$, with parameters $p$, $q$, where $q = e^{-\lambda}$ and $p = 1 - e^{-\lambda}$. It is known that the mean of $Z$ is equal to $q/p$ and the variance of $Z$ is equal to $q/p^2$. Hence, we have

$$E(Y) = E(Z) + 1 = \frac{e^{-\lambda}}{1 - e^{-\lambda}} + 1 = \frac{1}{1 - e^{-\lambda}} \text{ and} \quad (A2)$$

$$Var(Y) = Var(Z) = \frac{e^{-\lambda}}{\left(1 - e^{-\lambda}\right)^2}, \quad (A3)$$

which are equations (1) and (2), respectively. End of proof.

Proof of eq. (5) and (6): For convenience, we will define $f_1(x) = \lambda_1 e^{-\lambda_1 x}$ and $f_2(x) = \lambda_2 e^{-\lambda_2 x}$, so that $f(x) = \alpha_1 f_1(x) + \alpha_2 f_2(x)$. Let us introduce two random variables $X_1$ and $X_2$, such that $X_1$ has the pdf $f_1$ and $X_2$ has the pdf $f_2$. (We are not assuming that these two random variables are independent and are not implying that $X = X_1 + X_2$). We will also denote $Y_1 = \text{ceil}(X_1)$ and $Y_2 = \text{ceil}(X_2)$. For any $k \geq 1$, we compute

$$P(Y = k) = P(k-1 < X \leq k) = \int_{k-1}^{k} f(x)dx =$$

$$\alpha_1 \int_{k-1}^{k} f_1(x)dx + \alpha_2 \int_{k-1}^{k} f_2(x)dx =$$

$$\alpha_1 P(k-1 < X_1 \leq k) + \alpha_2 P(k-1 < X_2 \leq k) =$$

$$\alpha_1 P(Y_1 = k) + \alpha_2 P(Y_2 = k). \quad (A4)$$

We have already shown that $E(Y_i) = 1/1 - e^{-\lambda_i}$ and $Var(Y_i) = e^{-\lambda_i}/\left(1 - e^{-\lambda_i}\right)^2$, $i = 1,2$. Hence

$$E(Y) = \sum_{k=1}^{\infty} kP(Y = k) =$$

$$\alpha_1 \sum_{k=1}^{\infty} kP(Y_1 = k) + \alpha_2 \sum_{k=1}^{\infty} kP(Y_2 = k) = \alpha_1 E(Y_1) + \alpha_2 E(Y_2). \quad (A5)$$

Now, the expression for $E(Y)$, as stated in eq. (5), follows immediately from (A4). We can now find $E(Y)^2$. It is equal to

$$\sum_{k=1}^{\infty} k^2 P(Y = k) = \alpha_1 \sum_{k=1}^{\infty} k^2 P(Y_1 = k) + \alpha_2 \sum_{k=1}^{\infty} k^2 P(Y_2 = k) =$$

$$\alpha_1 E(Y_1^2) + \alpha_2 E(Y_2^2). \quad (A6)$$

Then eq. (6) for the variance of $Y$ can be immediately derived from (A5) and (A6) and the definition of variance. End of proof.

Proof of eq. (7) and (8): For any $k \geq 1$ we have

$$P(Y = k) = \int_{k-1}^{k} f(x)dx. \quad (A7)$$

Let us introduce $m(t)$, the moment generating function of $Y$,

$$m(t) = \sum_{k=1}^{\infty} e^{tk} P(Y = k) \quad (A8)$$

Then substituting (A7) into (A8) we obtain

$$m(t) = \sum_{k=1}^{\infty} e^{(t-\lambda)k}\left[\left(\lambda e^{\lambda} - \lambda\right)k - \left(\lambda e^{\lambda} - e^{\lambda} + 1\right)\right] \quad (A9)$$

which reduces to:

$$m(t) = \lambda\left(e^{\lambda} - 1\right)\frac{e^{t-\lambda}}{\left(1 - e^{t-\lambda}\right)^2} - \left(\lambda e^{\lambda} - e^{\lambda} + 1\right)\frac{e^{t-\lambda}}{1 - e^{t-\lambda}} \quad (A10)$$

The first moment of $Y$ is the first derivative of $m(t)$ evaluated at $t = 0$ and, similarly, the second moment of $Y$ is the second derivative of $m(t)$ evaluated at $t = 0$. From this we get eq. (7) and (8). End of proof.